\documentclass[preprint,superscriptaddress,nofootinbib]{revtex4-1}
\usepackage{graphicx, epsfig}

\usepackage{amsmath,amssymb}

\usepackage{color}

\begin{document}
\title{
Probing first-order electroweak phase transition via primordial black holes in the effective field theory}
%
\author{Katsuya~Hashino}
\affiliation{Department of Physics, Faculty of Science and Technology, Tokyo University of Science, Noda, Chiba 278-8510, Japan }
\author{Shinya~Kanemura}
\affiliation{Department of Physics, Osaka University, Toyonaka, Osaka 560-0043, Japan}
\author{Tomo~Takahashi}
\affiliation{Department of Physics, Saga University, Saga 840-8502, Japan}
\author{Masanori~Tanaka}
\affiliation{Department of Physics, Osaka University, Toyonaka, Osaka 560-0043, Japan}

\begin{abstract}

We investigate production of primordial black holes from first-order electroweak phase transition in the framework of the nearly aligned Higgs effective field theory, in which non-decoupling quantum effects are properly described. 
Since the mass of such primordial black holes is evaluated to be about $10^{-5}$ of the solar mass, current and future microlensing observations such as Subaru HSC, OGLE, PRIME and Roman Space Telescope may be able to probe the electroweak phase transition.
 We study parameter regions where primordial black holes can be produced by the first-order electroweak phase transition, and explore their detectability at these observations.
 Complementarity of primordial black hole observations, gravitational wave observations and collider experiments is also discussed for testing the nature of the electroweak phase transition.

 \end{abstract}

\preprint{OU-HET-1159}

\maketitle



\section{ Introduction }


Although a Higgs boson was discovered by the Large Hadron Collider (LHC) in 2012~\cite{ATLAS:2012yve,CMS:2012qbp}, dynamics of the electroweak symmetry breaking remains unknown and aspects of the electroweak phase transition (EWPT) in the early Universe is still a mystery.
In particular, the nature of the EWPT is essentially important to baryon asymmetry of the Universe which is one of the greatest problems in cosmology and particle physics. 
A scenario of electroweak baryogenesis~\cite{Kuzmin:1985mm} requires strongly first-order EWPT to satisfy 
the condition of departure from thermal equilibrium~\cite{Sakharov:1967dj}.
Since the strongly first-order EWPT cannot be realized in the standard model (SM)~\cite{Dine:1992vs,Kajantie:1995kf}, extensions of the SM are necessary for a successful scenario of electroweak baryogenesis.


As the extension from the SM for the strongly first-order EWPT, effective field theories have often been considered like the standard model effective field theory (SMEFT)~\cite{Grojean:2004xa, Delaunay:2007wb, Cai:2017tmh, Croon:2020cgk, Hashino:2021qoq}, the Higgs effective field theory (HEFT)~\cite{Banta:2022rwg} and its variations such as the nearly aligned Higgs effective field theory (naHEFT)~\cite{Kanemura:2022txx} etc.
In addition to the effective field theory approaches, studies in renormalizable models with an extended scalar sector have also been performed intensively~\cite{Turok:1991uc, Espinosa:1993bs, Funakubo:1993jg, Cottingham:1995cj, Cline:1996mga, Moreno:1996zm, Kanemura:2004ch, Funakubo:2005pu, Fromme:2006cm, Profumo:2007wc, Ahriche:2007jp, Noble:2007kk, Aoki:2008av, Funakubo:2009eg, Kanemura:2011fy, Espinosa:2011ax, Gil:2012ya, Fuyuto:2014yia, Tamarit:2014dua, Kanemura:2014cka, Profumo:2014opa, Blinov:2015vma, Fuyuto:2015vna, Karam:2015jta, Basler:2016obg, Dorsch:2017nza, Ghorbani:2017jls, Bernon:2017jgv, Chiang:2017nmu, Ghorbani:2019itr, Barman:2019oda}. 
In both approaches, it has turned out that there is a strong correlation between physics to satisfy the condition of strongly first-order EWPT and to make a large deviation from the SM prediction in the triple Higgs boson coupling ($hhh$ coupling)~\cite{Grojean:2004xa,Kanemura:2004ch}.
For example, the deviation larger than about 20$\%$ is predicted by the condition of strongly first-order EWPT in the two Higgs doublet model~\cite{Kanemura:2004ch, Enomoto:2021dkl, Kanemura:2022ozv}.
Such a large deviation in the $hhh$ coupling is expected to be measured at future collider experiments, such as High-Luminosity LHC (HL-LHC)~\cite{Cepeda:2019klc} and the International Linear Collider (ILC)~\cite{Asner:2013psa, Moortgat-Picka:2015yla, Fujii:2015jha}, by which a scenario of strongly first-order EWPT can be tested.


If the first-order phase transition occurs in the early Universe, gravitational waves (GWs) can be produced~\cite{Grojean:2006bp}.
The strongly first-order EWPT predicts specific form of GW spectrum, which is sensitive to the detailed shape of the Higgs potential of extended models from the SM.
The peak frequency of the GW spectrum from the EWPT is typically $10^{-3}$ -- $10^{-1}$Hz, which can be tested at the future space-based GW interferometers, such as Laser Interferometer Space Antenna (LISA)~\cite{Klein:2015hvg} and DECi-hertz Interferometer Gravitational Wave Observatory (DECIGO)~\cite{Yagi:2011wg}. 
 In recent years, many authors have studied complementarity of future collider experiments and future space-based GW observations~\cite{Kakizaki:2015wua,Hashino:2016rvx,Kobakhidze:2016mch,Huang:2016cjm,Hashino:2016xoj,Artymowski:2016tme,Beniwal:2017eik,Huang:2017rzf,Hashino:2018zsi,Chala:2018ari,Huang:2018aja,Bruggisser:2018mrt,Alves:2018oct,Hashino:2018wee,Ahriche:2018rao,Chala:2018opy,Alves:2018jsw,Alves:2019igs,Chen:2019ebq, Enomoto:2021dkl, Kanemura:2022txx, Kanemura:2022ozv}.


Recently, it has been discussed that the primordial black hole (PBH) observation can be used as a probe of strongly first-order EWPT~\cite{Hashino:2021qoq}, which is motivated by the work by Liu et. al.~\cite{Liu:2021svg} where PBH production and its abundances in general first-order phase transition have been studied. 
It has been known that large density contrast in the early Universe can gravitationally collapse into the PBHs~\cite{Hawking:1971ei,Carr:1974nx,Carr:1975qj}.
Sufficiently large density contrast can be generated by the first-order phase transition~\cite{Kodama:1982sf, Hawking:1982ga}, which can produce PBHs.
The mass of PBHs produced by the first-order EWPT is about $10^{-5}$ of the solar mass, which can be observed by microlensing observations such as Subaru Hyper Suprime Cam (HSC)~\cite{HSC}, Optical Gravitational Lensing Experiment (OGLE)~\cite{OGLE}, PRime-focus Infrared Microlensing Experiment (PRIME)~\cite{PRIME} and Nancy Grace Roman (Roman) Space Telescope~\cite{Roman}.
Therefore, exploration of such PBHs at these observations can be used to probe models of strongly first-order EWPT.


In the previous work~\cite{Hashino:2021qoq}, the authors evaluated the abundance of PBHs from first-order EWPT in the SMEFT.
Although large non-decoupling quantum effects of new particles in the fundamental model behind are important to realize strongly first-order EWPT, such effects may cause the SMEFT expansion to break down~\cite{Postma:2020toi,Kanemura:2022txx}.
In the present paper, we consider one of the realistic effective field theories, called the naHEFT~\cite{Kanemura:2021fvp} and investigate the production of PBHs in the framework.
This effective field theory can parameterize the non-decoupling quantum effects of the new physics at the next to leading order, by which strongly first-order EWPT can be well described.  
Therefore, the naHEFT is a useful framework to discuss the phenomena beyond the SM, such as the baryon asymmetry of the Universe. 
We study the parameter region in the naHEFT where PBHs can be produced from first-order EWPT and its detectability in the future observations.
 Complementarity of PBH observations, GW observations and collider experiments to probe the EWPT is also discussed.


In the next section, we first briefly review the naHEFT, and describe the phase transition parameters such as released latent heat and duration of phase transition in the framework.
We then clarify parameter region in which PBHs can be produced from the first-order EWPT.
In section III, we discuss the complementarity of PBHs, GWs and collider experiments to probe the strongly first-order EWPT.
We also examine that the parameter region, which is also related to extended Higgs models, can be widely explored by the PBH observations.
In the final section, we give conclusions and discussions.


 \section{Nearly aligned Higgs effective field theory}

Here, we briefly introduce the naHEFT~\cite{Kanemura:2021fvp,Kanemura:2022txx}. 
In this framework, the BSM part of the Higgs potential is given by 
\begin{align}
\label{BSM}
V_{\rm{BSM}}(\Phi) &= \frac{\xi}{4}\kappa_0\,[\mathcal{M}^2(\Phi)]^2 \ln\frac{\mathcal{M}^2(\Phi)}{\mu^2},
\end{align}
where $\Phi$ is the SM Higgs isospin doublet field, $\xi=1/(4\pi)^2$, $\kappa_0$ is a real dimensionless parameter, and $\mu^2$ is a real massive parameter.
The BSM effect to the Higgs potential is described by a Coleman-Weinberg form~\cite{Coleman:1973jx}.
The $\mathcal{M}^2(\Phi)$ in this equation is an arbitrary function of $|\Phi|^2$, which is assumed to take the following form  
\begin{align}
\label{mass}
\mathcal{M}^2(\Phi)\,=\,M^2 + \kappa_{\rm{p}}\, |\Phi|^2,
\end{align}
where $M^2$ and $\kappa_{\rm{p}}$ are real parameters.
In order to discuss the non-decoupling effect, we introduce parameters $\Lambda$ and $r$, which are given by 
\begin{align}
\label{parameters}
\Lambda^2 &=M^2 + \frac{\kappa_{\rm{p}}}{2}v^2,\\
r&=\frac{\frac{\kappa_{\rm{p}}}{2}v^2}{\Lambda^2}=1-\frac{M^2}{\Lambda^2}.
\end{align}
The dimensionful parameter $\Lambda$ corresponds to the physical mass of the integrated new particles. 
The parameter $r$ shows non-decoulingness of the new physics behind.
When $r\to0$ ($r\to1$), $\Lambda\sim \sqrt
{M^2}$ ($\Lambda\sim \sqrt{\kappa_pv^2/2}$). 
Thus, the quantum effects for non-decoupling in the potential are controlled by the parameter $r$.
 Therefore, the scalar sector of the naHEFT has three independent parameters:
\begin{align}
\label{freeparameters}
\kappa_0,\quad \Lambda,\quad  r.
\end{align}
For example, if the naHEFT is realized by integrating out additional scalar bosons, whose masses are degenerated, in the model with an extended Higgs sector, $\kappa_0$ and $\Lambda$ correspond to the number of additional scalar fields and the physical mass of additional scalar bosons, respectively.
We explore the parameter region where PBHs can be produced from first-order EWPT.

The effective potential in the naHEFT is given at the temperature $T$ by~\cite{Kanemura:2022txx}
\begin{align}
\label{eq:BSMeffectiveV}
V_{\rm{eff}}(\varphi,T) \,=\, V^{\rm{SM}}_{\rm{eff}}(\varphi,T) + V_{\rm{BSM}}(\varphi) \,+\,\Delta V_{{\rm BSM},T}(\varphi,T)\,,
\end{align}
where $V^{\rm{SM}}_{\rm{eff}}$ is the effective potential in the SM and $\varphi$ is the classical scalar field with $\langle\Phi \rangle=(0,\varphi/\sqrt{2})^T$.
Here, the finite temperature contributions coming from the BSM effects are given by
\begin{align}
\Delta V_{{\rm BSM},T}(\varphi,T)&\,=\, 8\,\xi\, T^4\, \kappa_0\,J_{\rm{BSM}}\left(\frac{\mathcal{M}^2(\varphi)}{T^2}\right)\,.
\end{align}
The $J_{\rm{BSM}}$ function is given by
\begin{align}
J_{\rm{BSM}}\left(\frac{\mathcal{M}^2(\varphi)}{T^2}\right)\,=\, \int^\infty_0 dk^2 k^2 \ln\left( 1-\mbox{sign}(\kappa_0)\, e^{-\sqrt{k^2+\frac{\mathcal{M}^2(\varphi)}{T^2}}} \right) \,,
\end{align}
where $\mbox{sign}(\kappa_0)$ is positive (negative) in the case of $\kappa_0>0$ $(\kappa_0<0)$.

\begin{figure}[t]
  \begin{center}
\includegraphics[width=0.4\textwidth]{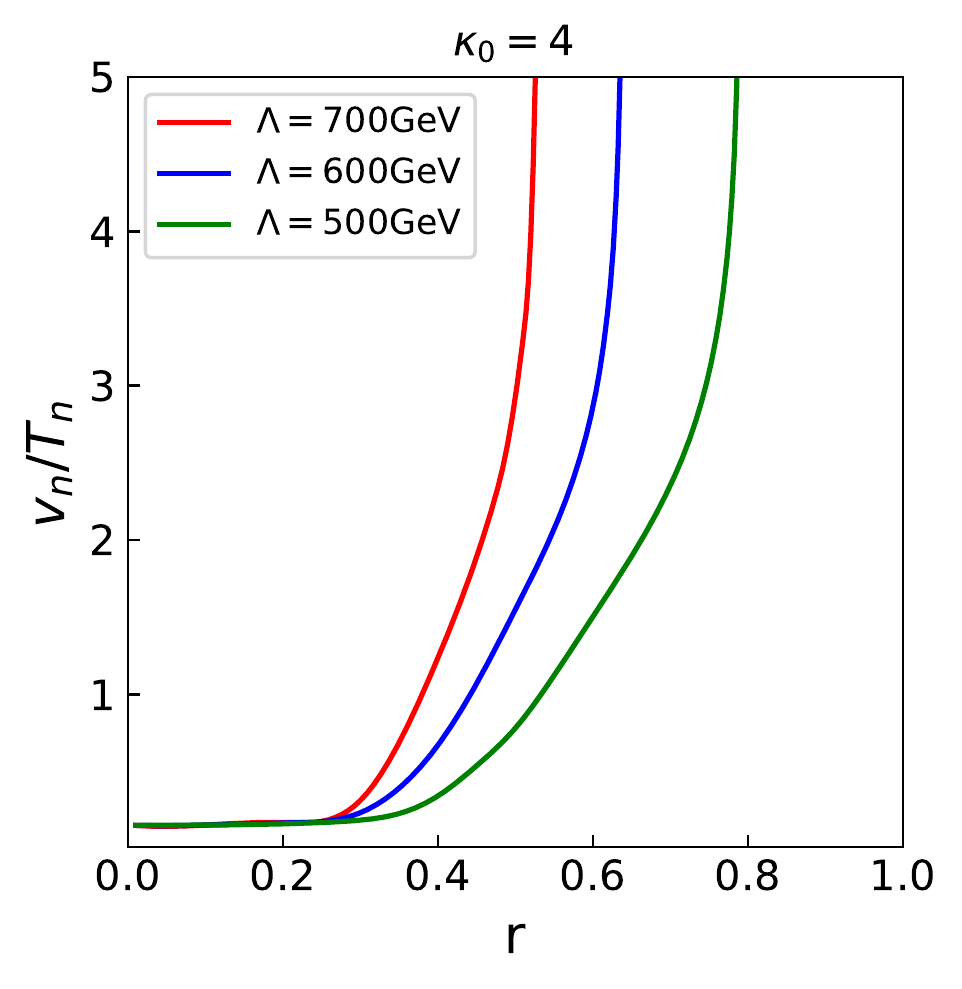}
\includegraphics[width=0.4\textwidth]{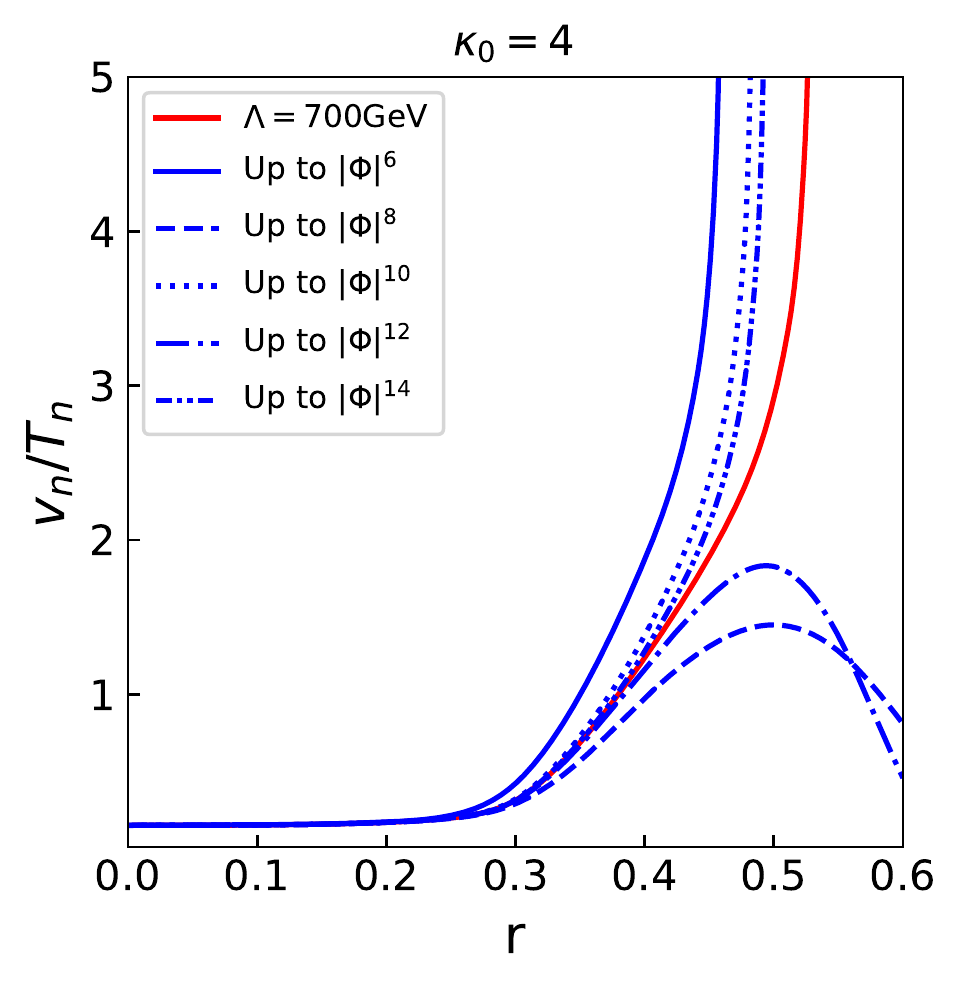}
\caption{ (Left) Value of $v_n/T_n$ in the naHEFT with $\kappa_0=4$ and $\Lambda=500$, 600 and 700 GeV.
The green, blue and red lines correspond to $\Lambda=500$, 600 and 700 GeV, respectively. 
(Right) The value of $v_n/T_n$ in the naHEFT and the SMEFT approximation with $\Lambda=$ 700 GeV.
The blue solid, blue dashed, blue dotted, blue dot-dashed and blue dashed double-dotted lines represent the results in the potential of naHEFT truncated up to $|\Phi|^6$, $|\Phi|^8$, $|\Phi|^{10}$, $|\Phi|^{12}$ and $|\Phi|^{14}$, respectively.
}
\label{phinTn}
  \end{center}
\end{figure}
Due to the effects of $V_{\rm eff}^{\rm BSM}$, the strongly first-order EWPT, which is represented by the condition of $v_n/T_n>1$, can be realized.
Here, $v_n$ is the true vacuum at the nucleation temperature $T_n$, at which one bubble nucleates in the Hubble volume.
$T_n$ is computed by calculating the bounce solution of equation of motion~\cite{Linde:1981zj}. 
Numerical results of $v_n/T_n$ in the naHEFT with $\kappa_0=4$ are shown in the left panel of Fig.~\ref{phinTn}.
The horizontal and vertical axises of this figure are the parameter $r$ and the value of $v_n/T_n$, respectively.
The green, blue and red lines in the left panel correspond to the $v_n/T_n$ values of the naHEFT with $\Lambda=500$, 600 and 700 GeV, respectively.
Curvatures of these lines change when the parameter $r$ is relatively large, since the cubic term $\varphi^3$ coming from finite temperature effects is no longer important in generating a sizable barrier to realize the first-order EWPT.
The right panel shows the results of $v_n/T_n$ in the potential of naHEFT and ones truncated up to $|\Phi|^6$, $|\Phi|^8$, $|\Phi|^{10}$, $|\Phi|^{12}$ and $|\Phi|^{14}$.
The red solid line in the right panel is the value of $v_{n}/T_{n}$ in the naHEFT with $\kappa_{0} = 4$ and $\Lambda = 700$ GeV.
The blue solid, blue dashed, blue dotted, blue dot-dashed and blue dashed double-dotted lines in the right panel are the value of $v_n/T_n$ in the potential with $\kappa_0=4$ and $\Lambda=700$ GeV truncated up to $|\Phi|^6$, $|\Phi|^8$, $|\Phi|^{10}$, $|\Phi|^{12}$ and $|\Phi|^{14}$, respectively.
At relatively large values of $r$, $v_n/T_n$ in the naHEFT rapidly blows up, since the temperature for starting the EWPT gets close to zero.
In the case of relatively large $r$, deviations on $v_n/T_n$ for the SMEFT with truncation up to $|\Phi|^6$ --  $|\Phi|^{14}$ operators from the prediction in the naHEFT become large.
This implies that the naHEFT is better than the SMEFT to discuss the parameter region for strongly first-order EWPT.

It is well known that in the parameter region where the strongly first-order EWPT is realized a large deviation is predicted in the $hhh$ coupling from the SM value, which is given by 
	\begin{align} 
	\frac{\Delta\lambda_{hhh}}{\lambda_{hhh}^{\rm SM}} \equiv \frac{\lambda_{hhh}-\lambda_{hhh}^{\rm SM}}{\lambda_{hhh}^{\rm SM}},\quad \lambda_{hhh}= \left.\frac{\partial^3 V_{\rm eff}(\varphi)}{\partial \varphi^3}\right|_{\varphi=v},
	\end{align}
where effective potential $V_{\rm eff}(\varphi,0)$ is given in Eq.~(\ref{eq:BSMeffectiveV}), which contains one-loop level quantum corrections\footnote{ 
 One-loop corrections to the $hhh$ coupling in extended Higgs models is discussed in Ref.~\cite{Kanemura:2004mg,Aoki:2012jj,Arhrib:2015hoa,Hashino:2015nxa,Kanemura:2016lkz}.
Typically, the two-loop effects give positive contributions to the $hhh$ coupling, which are about 20$\%$ of those at one-loop~\cite{Braathen:2019pxr,Braathen:2019zoh,Braathen:2020vwo}.}.  
Therefore, the strongly first-order EWPT can be tested by precisely measuring the $hhh$ coupling at future collider experiments~\cite{Kanemura:2004ch}.  
At the HL-LHC, the $hhh$ coupling can be measured at 50 $\%$ accuracy (at the 68$\%$ confidence level)~\cite{Cepeda:2019klc}, while at the future lepton collider such as the ILC energy upgraded version with the energy 500 GeV and 1 TeV, the $hhh$ coupling is expected to be measured by 27$\%$ and 10$\%$ accuracies (at the 68$\%$ confidence level), respectively~\cite{Bambade:2019fyw}.

 \section{ First-order EWPT and PBH in naHEFT}

We here introduce phase transition parameters $\alpha$ and $\beta/H$.
$\alpha$ is the normalized released latent heat by radiative energy density, which is defined as
	\begin{align} 
	\alpha\equiv \epsilon(T_n)/ \rho_{\rm rad}(T_n),
	\label{alphaGW}
	\end{align}
where the $\rho_{\rm rad}(T)=(\pi^2/30)g_* T^4$ with $g_*=106.75$, and $\epsilon$ is given by
 \begin{align}
 \label{latenth}
  \epsilon(T)
  =  \Delta V_{\rm eff} -T
  \frac{\partial  \Delta V_{\rm eff} }{\partial T},\quad  \Delta V_{\rm eff} =  V_{\rm eff}(\varphi_-(T),T) - V_{\rm eff}(\varphi_+(T),T),
\end{align}
where $\varphi_{+}$ and $\varphi_{-}$ denote the true and false vacua, respectively.
The $\beta/H$ corresponds to the inverse of the duration of phase transition, and is defined as 
	\begin{align} 
	\frac{\beta}{H}\equiv T_n\left.\frac{d}{dT}\left(\frac{S_3}{T}\right)\right|_{T=T_n},
	\label{betaGW}
	\end{align}
where $S_3$ is the three-dimensional Euclidian action.

If the first-order phase transition occurs in the early Universe, GWs can be produced due to the bubble dynamics of the true vacuum. 
The GWs from the first-order EWPT have three sources: collisions of the vacuum bubbles, compressional waves (sound waves) and magnetohydrodynamics turbulence.
The leading contribution among the three sources comes from the sound waves, whose amplitude is given by~\cite{Caprini:2015zlo}
\begin{align}
  \Omega_{\rm sw}(f)h^2
  = 2.65 \times 10^{-6} v_{w} \left( \frac{H}{\beta} \right) \left( \frac{\kappa_{v} \alpha }{1+\alpha} \right)^2 \left( \frac{100}{g_{*}}\right)^{1/3} (f/f_{\rm sw})^3 \left( \frac{7}{4+3(f/f_{\rm sw})^2} \right)^{7/2},
\end{align} 
where $v_{w}$ is the wall velocity, and $\kappa_{v}$ is the fraction of the released latent heat contributing to sound wave formation, which is given in Ref.~\cite{Espinosa:2010hh}.
The peak frequency $f_{\rm sw}$ is given by~\cite{Caprini:2015zlo}
\begin{align}
  f_{\rm sw} = 1.9 \times 10^{-2} {\rm mHz} \frac{1}{v_{w}} \left( \frac{\beta}{H}  \right) \left( \frac{T_{n}}{100 {\rm GeV}} \right) \left( \frac{g_{*}}{100}\right)^{1/6}.
\end{align}
The prediction of the GW spectrum in the naHEFT was investigated in Ref.~\cite{Kanemura:2022txx}. 
To examine parameter regions where GWs can be detected at LISA and DECIGO, we use the signal-to-noise ratio for the observation of the GW spectrum, which is discussed in Ref.~\cite{Cline:2021iff}. 
The criterion is such that the signal-to-noise ratio for the GW spectrum is larger than ten, which is adopted in our later analysis.

Next, it is known that sufficient large density contrast in the early Universe results in overdensity regions which can gravitationally collapse into PBHs.
According to Ref.~\cite{Liu:2021svg}, the large density contrast can be generated by delaying the vacuum bubble nucleation.
Since the vacuum bubble nucleation is probabilistic, there is a possibility that the symmetry breaking is delayed in a whole Hubble volume.
The vacuum energy density in the unbroken symmetry region is larger than that in the broken symmetry region because the difference of the vacuum energy density is related to the difference of the height of the Higgs potential.
This energy density difference leads to the energy density contrast between the inside and outside of the Hubble volume in which the symmetry breaking is delayed.
The energy density contrast is defined as
\begin{align}
\label{cont}
\delta \equiv \frac{ | \rho_{\rm in} - \rho_{\rm out}|}{\rho_{\rm out}},
\end{align}
where $\rho_{\rm in}$ and $\rho_{\rm out}$ is the total energy density inside and outside the Hubble volume, respectively.
When the energy density contrast $\delta$ exceeds the critical value $\delta_{c} = 0.45$~\cite{Musco:2004ak,Harada:2013epa}, the inside of the Hubble volume can gravitationally collapse into a PBH.
Therefore, the mass of PBHs is roughly determined by the Hubble horizon mass at the time when the PBHs are produced. 
For the EWPT, the mass of PBHs is about $10^{-5} M_{\odot}$ ($M_{\odot}$ is the solar mass).
Thus, the fraction of the PBHs $f_{\rm PBH}$ produced from the first-order EWPT can be probed by the current and future microlensing observations such as Subaru HSC, OGLE, PRIME and Roman Space Telescope.
The fraction $f_{\rm PBH}$ can be determined by the phase transition parameters $\alpha$ and $\beta/H$ in Eqs.~\eqref{alphaGW} and \eqref{betaGW}. 
The method of calculating the fraction $f_{\rm PBH}$ is explained in the Appendix.

\begin{figure}[t]
  \begin{center}
\includegraphics[width=0.4\textwidth]{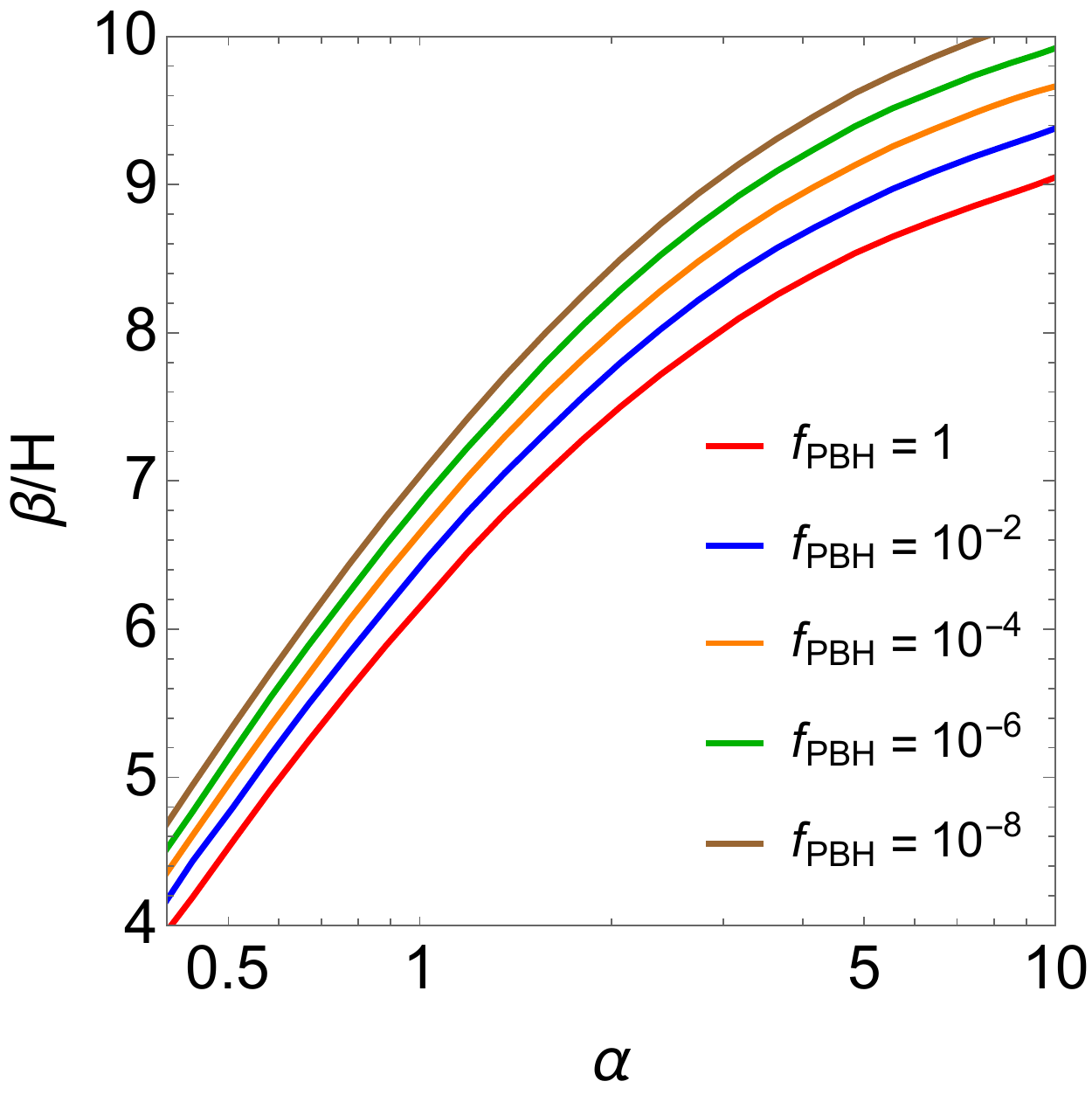}
\caption{ Contours for the PBH abundance are shown in the $\alpha$ and $\beta/H$ plane.
Red, blue, orange, green and brown lines correspond to $f_{\rm PBH} = 1, 10^{-2}, 10^{-4}, 10^{-6}$ and $10^{-8}$, respectively.
The region $10^{-4}<f_{\rm PBH}<1$, between red and orange lines, can be explored by PBH observations, such as PRIME and Roman Space Telescope.
 In the white region above the brown line, the abundance of the PBH cannot be produced or otherwise is too small.
 In the white region below the red line, PBHs are overproduced ($f_{\rm PBH}>1$). }
\label{PBH}
  \end{center}
\end{figure}

Fig.~\ref{PBH} represents model independent numerical results of $f_{\rm PBH}$ in the $\alpha$-$\beta/H$ plane, which were discussed in Ref.~\cite{Hashino:2021qoq}.
The red, blue, orange, green and brown lines correspond to $f_{\rm PBH} = 1, 10^{-2}, 10^{-4}, 10^{-6}$ and $10^{-8}$, respectively.
Current microlensing experiments, such as Subaru HSC and OGLE, can explore the region between the red and blue lines with $10^{-2}<f_{\rm PBH}<1$.
On the other hand, the region between the red and orange lines with $10^{-4}<f_{\rm PBH}<1$ can be explored by future microlensing experiments such as PRIME and Roman Space Telescope. 
In the white region above the brown line, the PBH abundance becomes too small or the PBHs cannot be produced from the first-order EWPT.
In the white region below the red line, PBHs are overproduced ($f_{\rm PBH}>1$).
According to this figure, we can discuss whether PBH observations can be used to probe the strongly first-order EWPT.

\begin{figure}[t]
  \begin{center}
\includegraphics[width=0.43\textwidth]{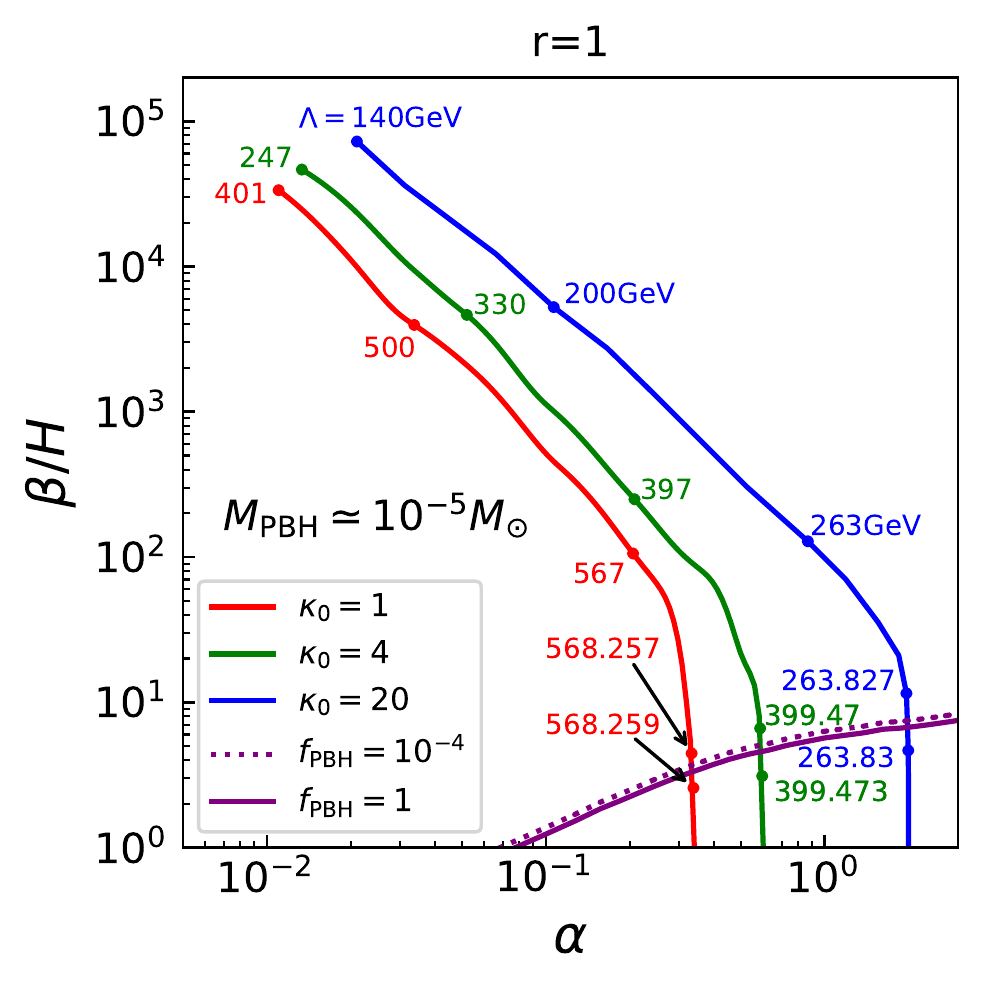}
\includegraphics[width=0.43\textwidth]{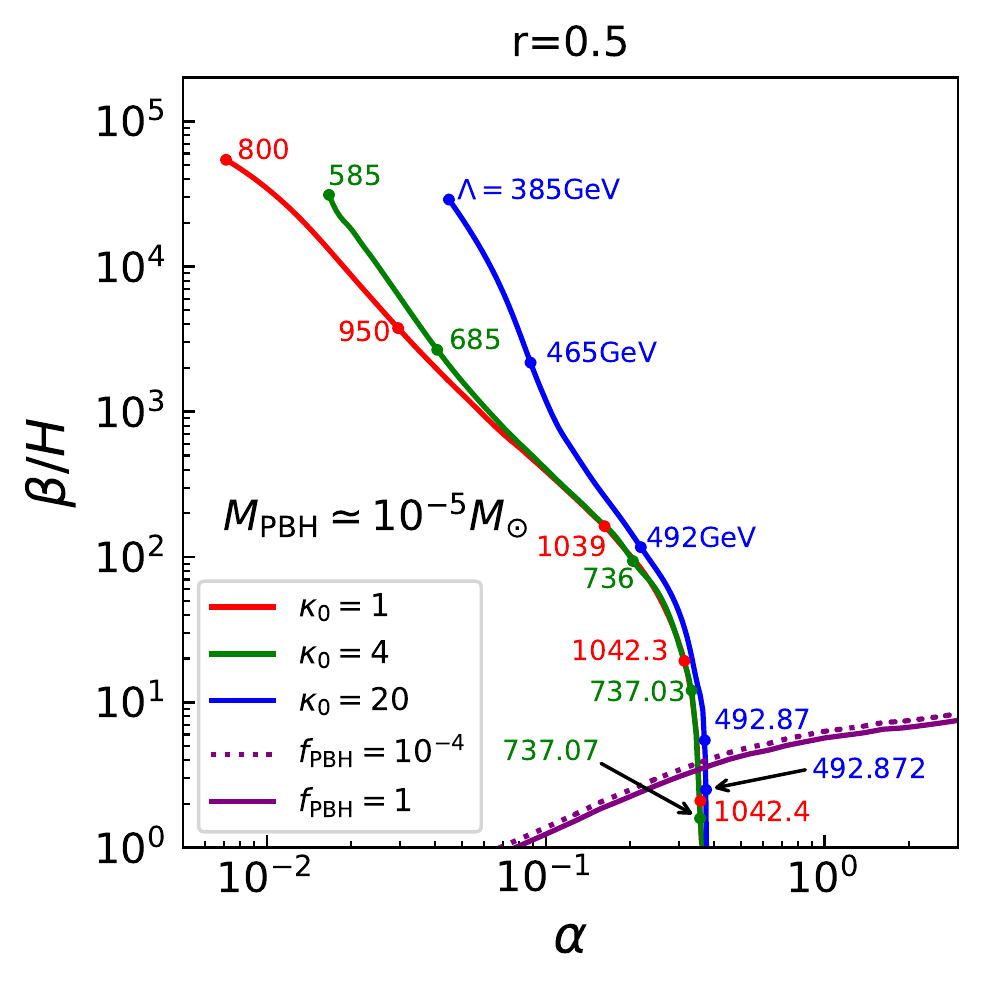}
\caption{ Parameters $\alpha$ and $\beta/H$ with respect to $\kappa_0$, $\Lambda$ and $r$.
Left (right) figure represents $\alpha$ and $\beta/H$ in the naHEFT with $r=1$ (0.5).
The red, green and blue lines are $\kappa_0=1$, 4 and 20, respectively.
The points on these lines correspond to the $\Lambda$ value.
Purple dotted and solid lines respectively are $f_{\rm PBH} = 10^{-4}$ and 1.}
\label{abH}
  \end{center}
\end{figure}
Fig.~\ref{abH} represents the parameters $\alpha$ and $\beta/H$ in the naHEFT.
The parameter $r$ is assumed as $r$ =1 and 0.5 in the left and right panels, respectively.
Red, green and blue lines correspond to the parameters $\alpha$ and $\beta/H$ with $\kappa_0=$ 1, 4 and 20, respectively.
Points on these lines in this figure represent the value of $\Lambda$. 
Purple dotted and solid lines respectively correspond to $f_{\rm PBH} = 10^{-4}$ and 1.
We here comment on the behaviors of the lines in right panel of Fig.~\ref{abH}, which are degenerate in the case of large $\Lambda$ value. 
The behavior is ascribed to the Boltzmann suppression with respect to finite temperature effects:  $\Delta V_{{\rm BSM},T}(0,T)  \propto \exp \left[ -  \Lambda^2 (1-r)/T^2 \right]$.
For small $r$ and large $\Lambda$, the potential is mainly determined by the zero temperature BSM effects.
Then the difference between the origin and the bottom of the potential, which is related to the phase transition parameters, is roughly given by $\kappa_0 r \Lambda^4$. 
For example, cases with ($\kappa_0$, $\Lambda$ [GeV], $r$) = (1, 1039, 0.5), (4, 736, 0.5) and (20, 492, 0.5), which are depicted in the right panel of Fig.~\ref{abH}, have almost the same value of $\kappa_0 r \Lambda^4$, and thus these points get close to each other actually. 

%
%

\begin{figure}[t]
  \begin{center}
\includegraphics[width=0.4\textwidth]{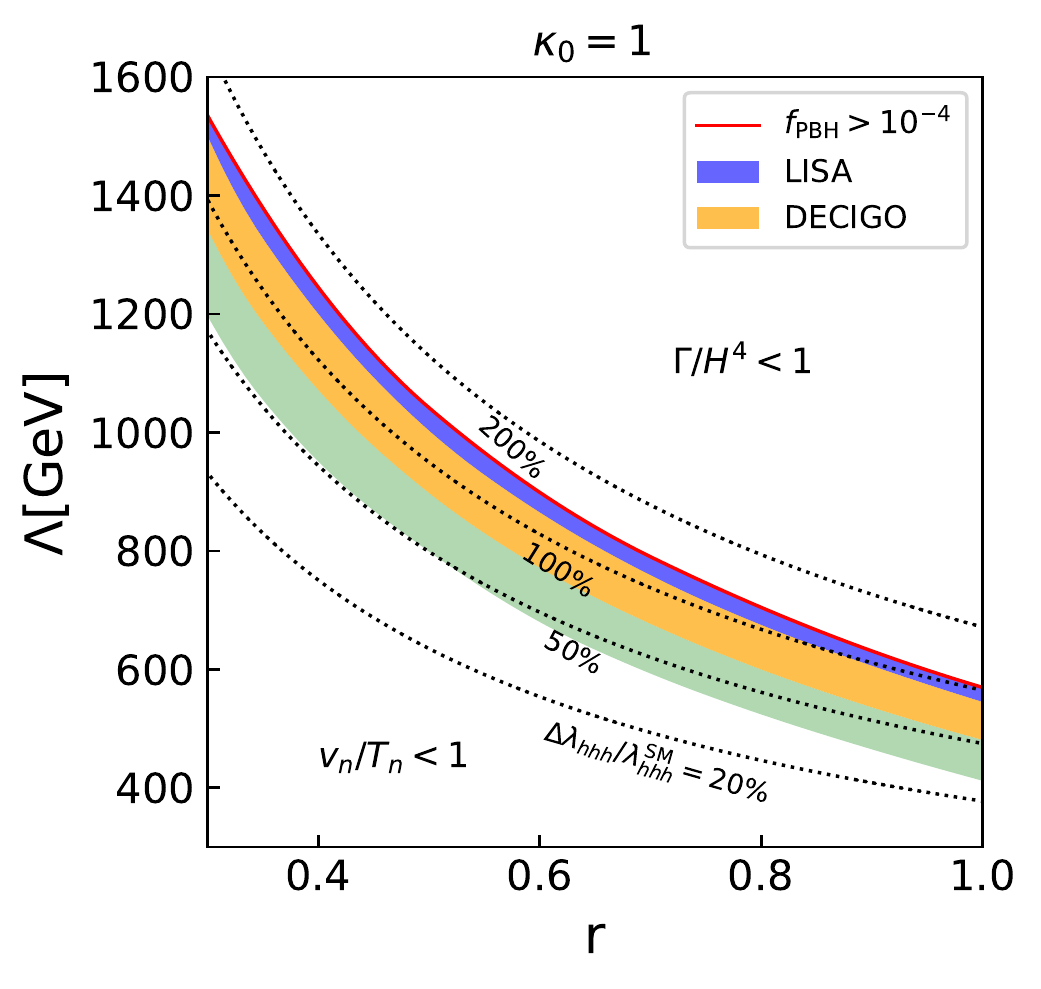}
  \includegraphics[width=0.4\textwidth]{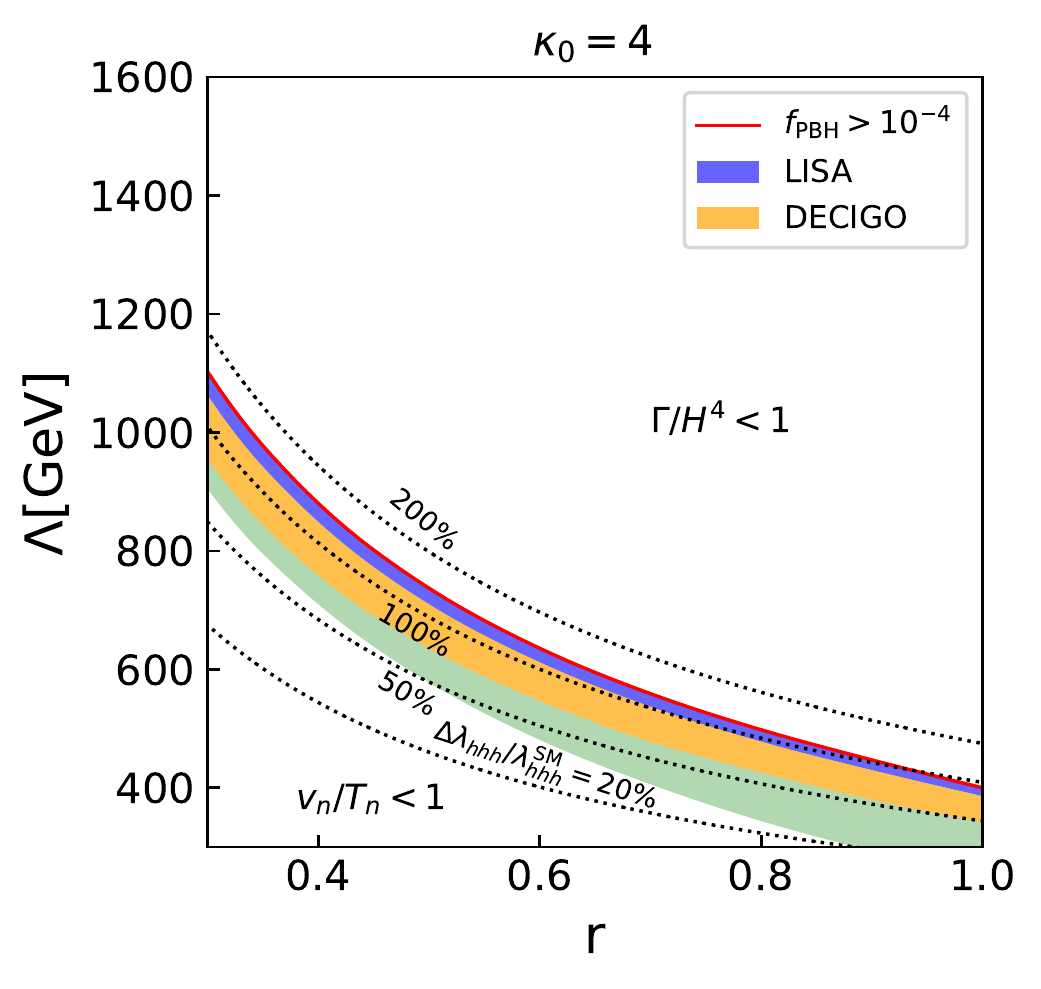}
\includegraphics[width=0.4\textwidth]{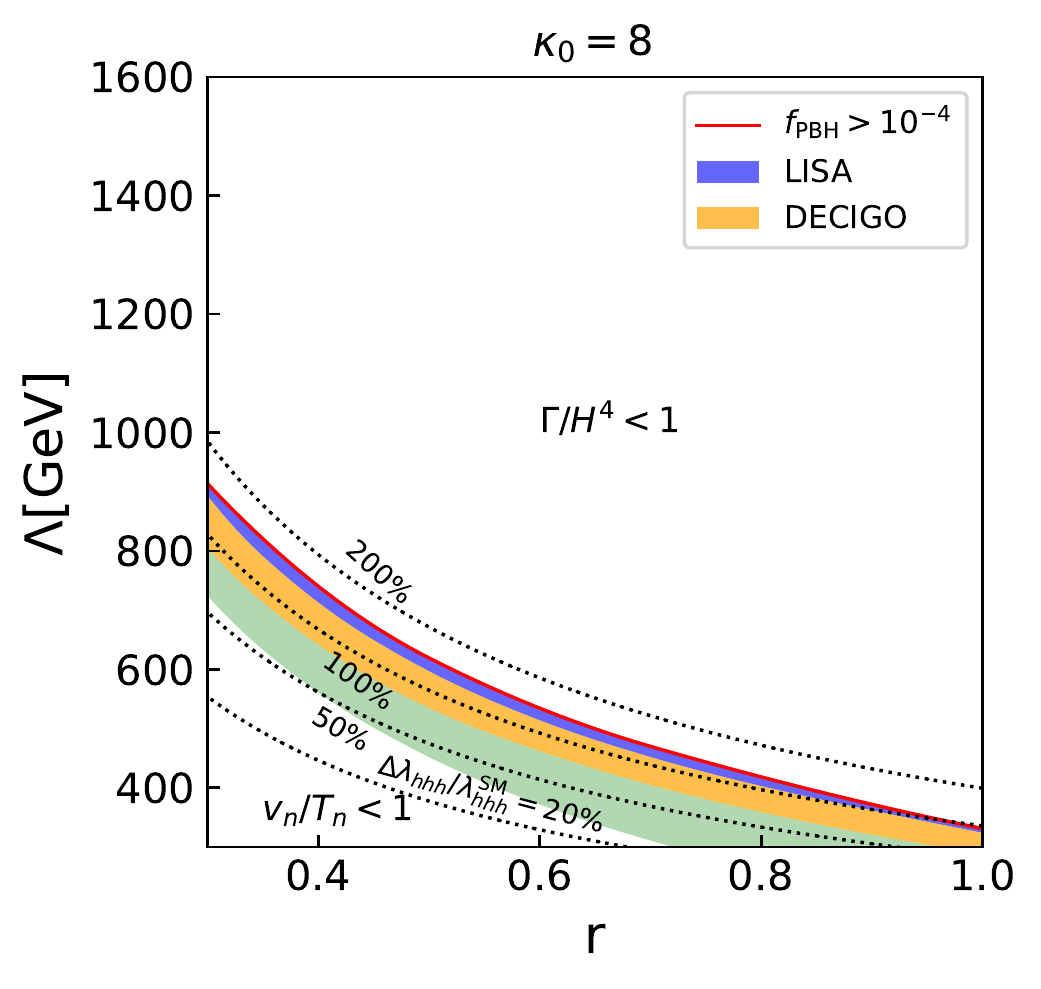}
\includegraphics[width=0.4\textwidth]{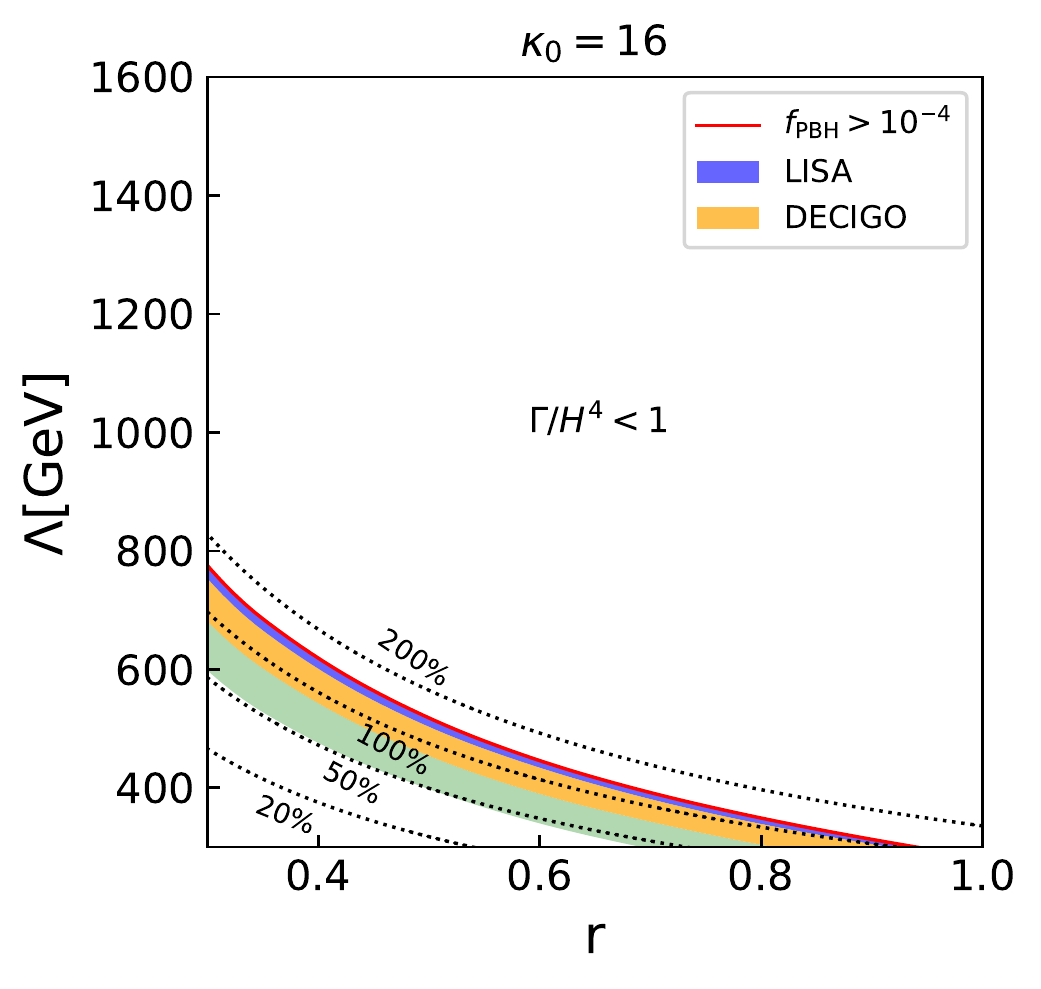}
\caption{ 
Regions of strongly first-order EWPT, where $v_n/T_n\geq1$, are shown as colored regions in the $r$-$\Lambda$ plane for $\kappa_0=1$, 4, 8 and 16.
In the red region, $f_{\rm PBH}$ can be larger than $10^{-4}$.
The EWPT has not been finished at the current Universe in top right white regions above the red one: $\Gamma/H^4<1$.
The orange regions represent that the detectable GW at DECIGO experiment can be produced.
The GW spectrum for the blue and red regions can be observed by both LISA and DECIGO experiments. 
The black dotted lines are the deviation in the $hhh$ coupling from the SM prediction value $\Delta\lambda_{hhh}/\lambda_{hhh}^{\rm SM}$ = 20, 50, 100 and 200 $\%$ from the bottom, respectively.
}
\label{abH2}
  \end{center}
\end{figure}

Fig.~\ref{abH2} represents the region of strongly first-order EWPT in the naHEFT with $\kappa_0=$ 1, 4, 8 and 16 in the $r$-$\Lambda$ plane.
For example, assuming the O(N) singlet scalar field theory as the UV theory, the parameter region with $r<0.3$ would be prohibited by the perturbative unitarity bound~\cite{Hashino:2016rvx}, and thus, we do not take into account such parameter regions in the following numerical analysis.
The black dotted lines represent $\Delta\lambda_{hhh}/\lambda_{hhh}^{\rm SM}$ = 20, 50, 100 and 200 $\%$ from the bottom, respectively. 
In the red region, the $f_{\rm PBH}$ can be larger than $10^{-4}$.
Thus, the first-order EWPT can be tested using the PBH observations.
In this region, we can also use GWs to test the EWPT. 
In the blue (orange) parameter region, the first-order EWPT can be tested by using GW observation at both LISA and DECIGO (only at DECIGO). 
In the green region, although GWs cannot be detected at LISA nor DECIGO the first-order EWPT can still be tested by the precision measurement of the $hhh$ coupling at future collider experiments. 
In the top right white region above the red solid line of these panels, the EWPT has not been completed at the current Universe, in which $\Gamma/H^4<1$.
In the bottom left white region below the colored region of this figure, the strongly first-order EWPT cannot be realized because of $\varphi_C/T_C<1$.

\begin{figure}[t]
  \begin{center}
\includegraphics[width=0.4\textwidth]{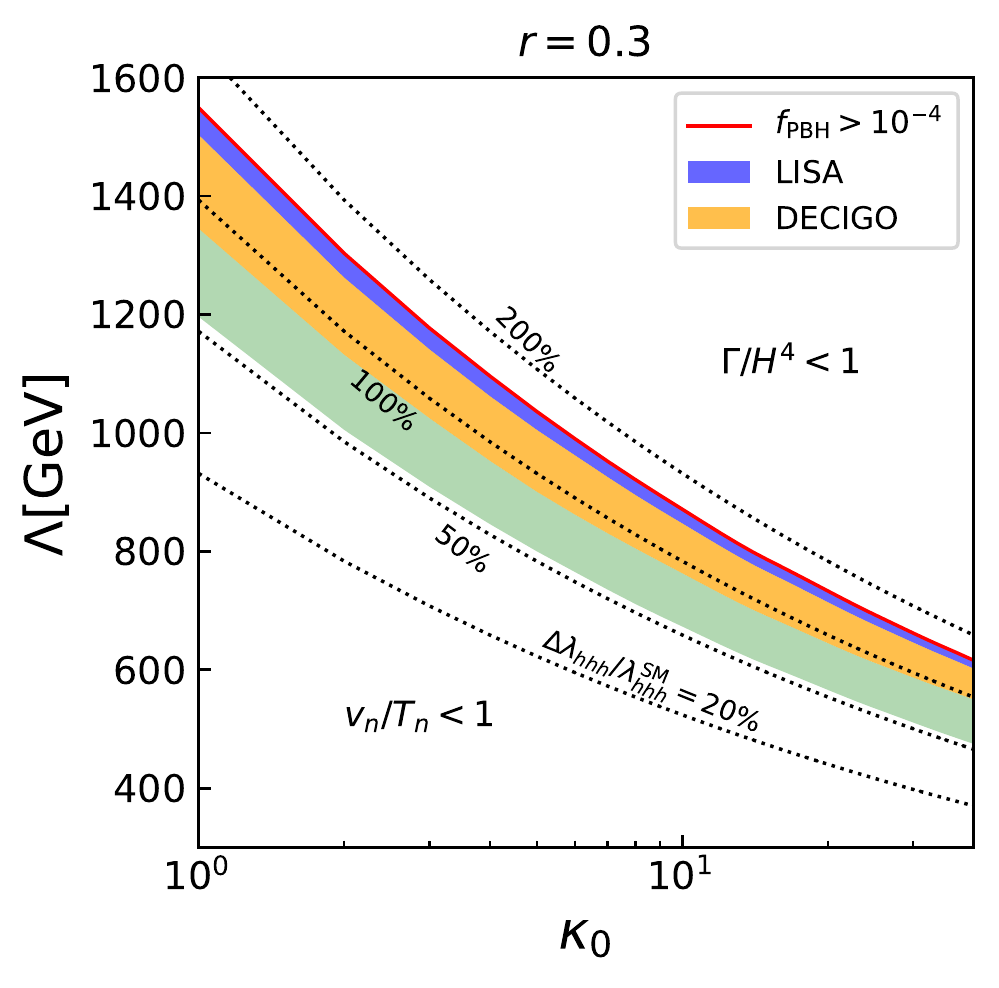}
\includegraphics[width=0.4\textwidth]{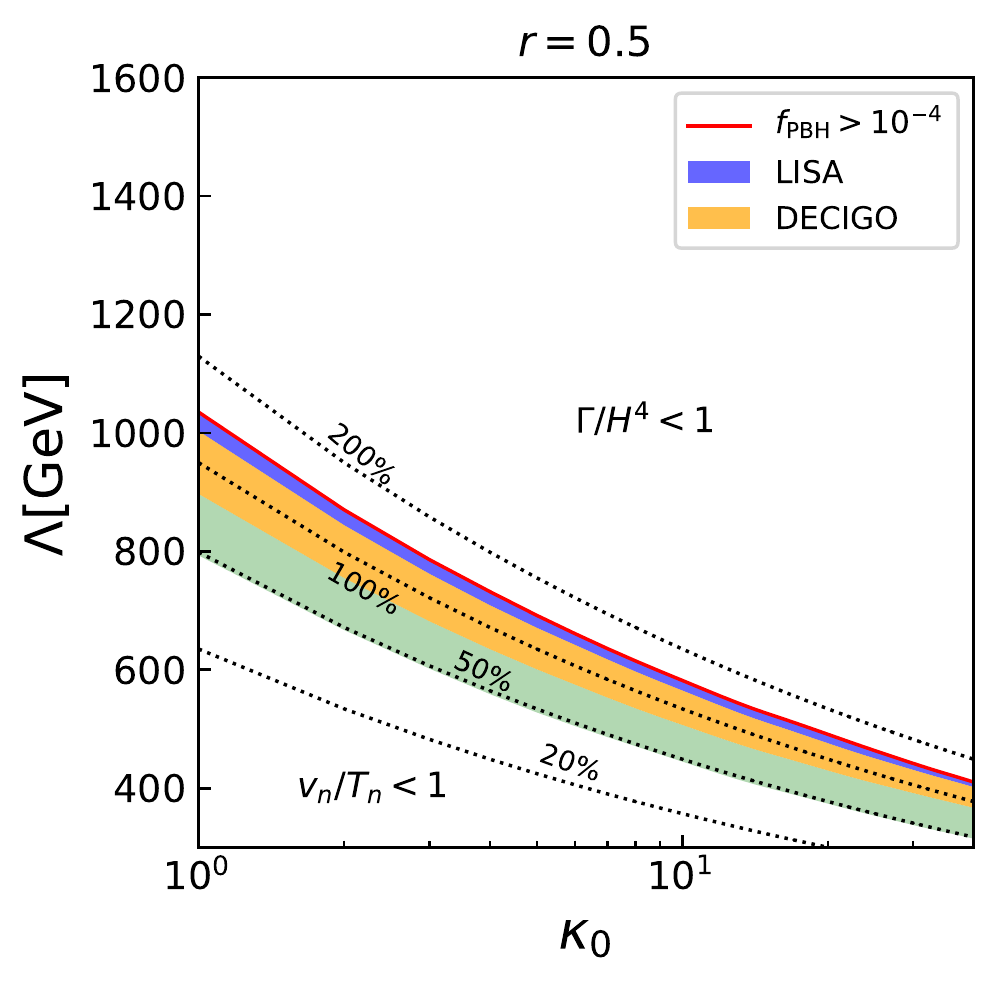}
\includegraphics[width=0.4\textwidth]{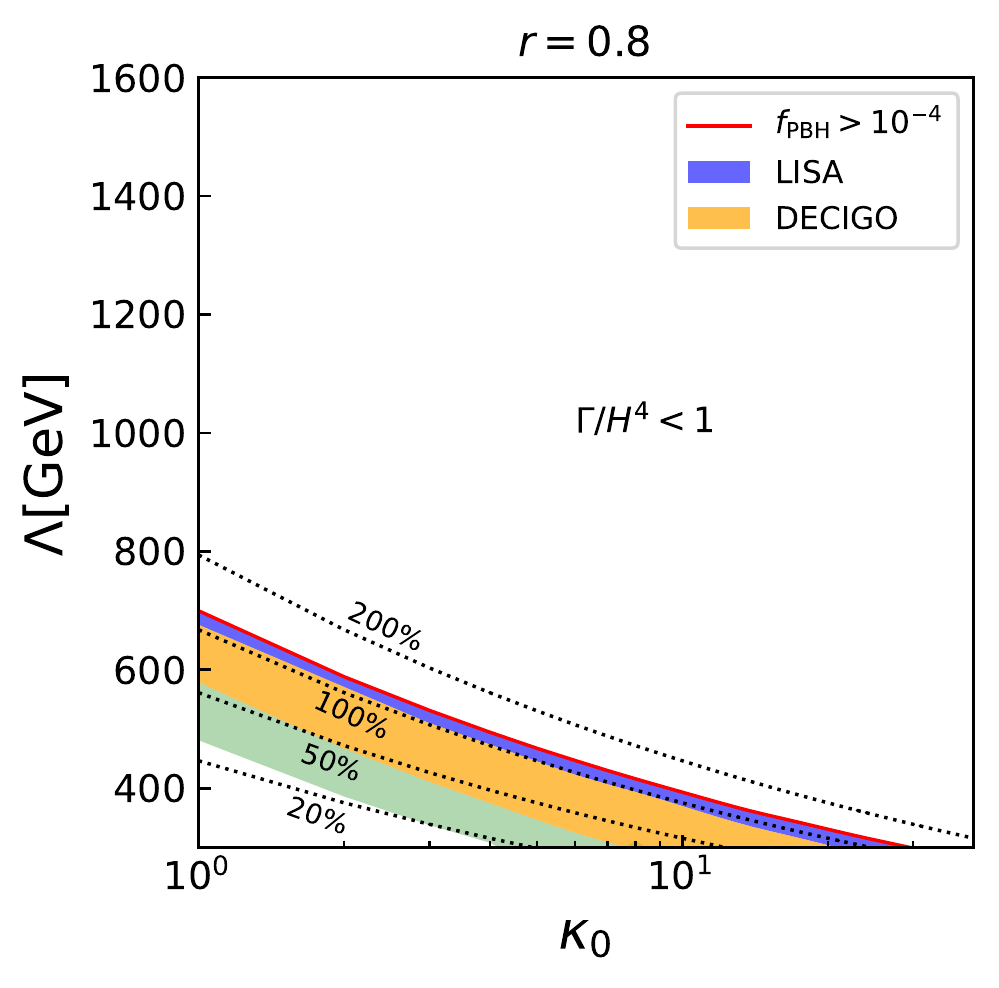}
\includegraphics[width=0.4\textwidth]{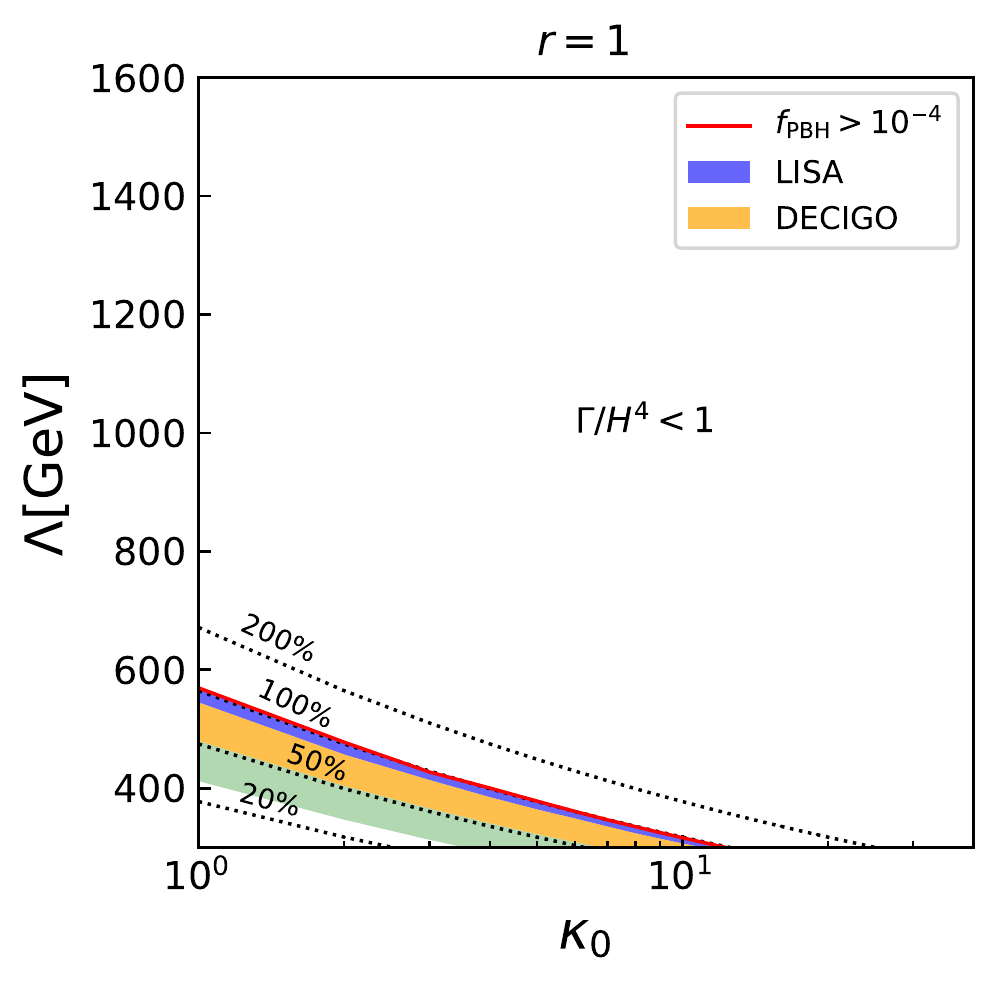}
\caption{  
Regions of strongly first-order EWPT, where $v_n/T_n\geq1$, are shown as colored regions in the $\kappa_0$-$\Lambda$ plane for $r=0.3$, 0.5, 0.8 and 1.
Otherwise the same as Fig.~\ref{abH2}.
}
\label{abH3}
  \end{center}
\end{figure}

Fig.~\ref{abH3} shows the region of strongly first-order EWPT in the naHEFT with $r=0.3$, 0.5, 0.8 and 1 in the $\kappa_0$ --$\Lambda$ plane. 
The colored regions and black dotted contours have the same definitions in Fig.~\ref{abH2}. 
For large $\kappa_0$ value, the strongly first-order EWPT can be realized by small $\Lambda$ value. 
According to these Figs.~\ref{abH2} and \ref{abH3}, the naHEFT can be complementarily tested by collider experiments, GW and PBH observations.

\begin{figure}[t]
  \begin{center}
\includegraphics[width=0.45\textwidth]{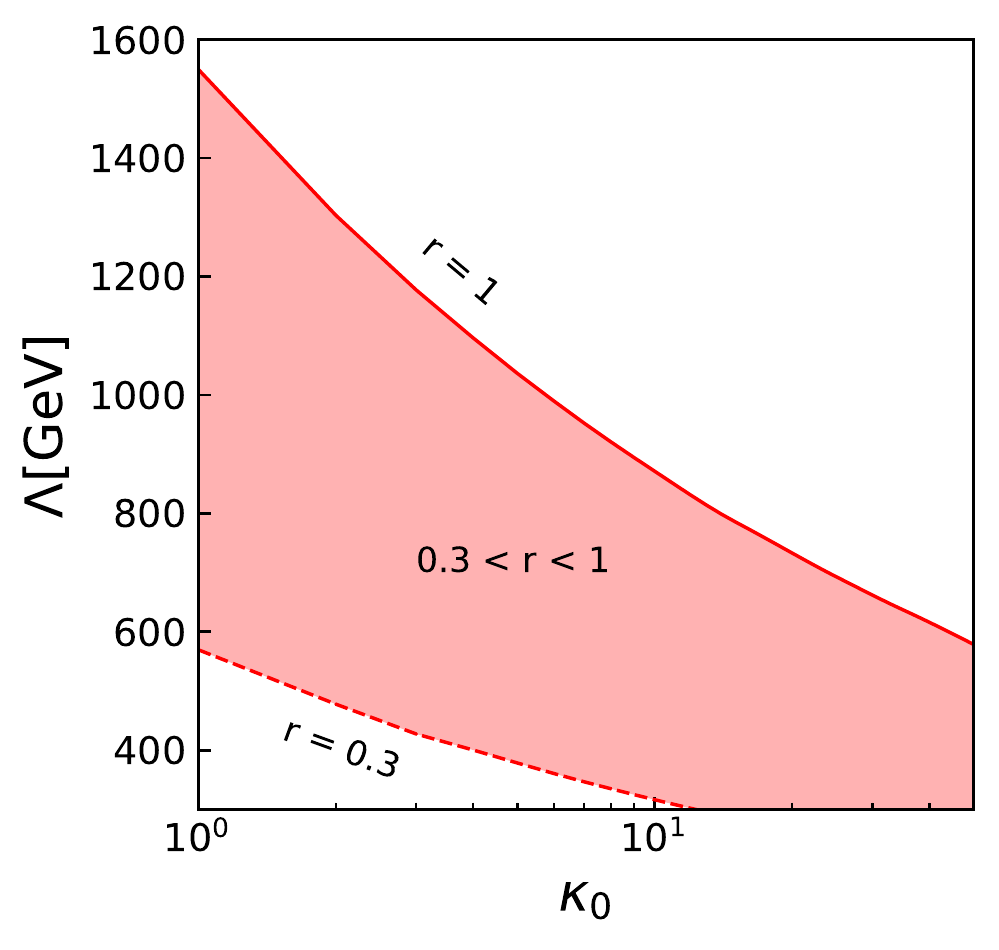}
\caption{  
The parameter region where PBHs from strongly first-order EWPT may be able to be detected in the $\kappa_0$-$\Lambda$ plane ($f_{\rm PBH}>10^{-4}$).
}
\label{abH4}
  \end{center}
\end{figure}

Fig.~\ref{abH4} shows the parameter region where PBHs from strongly first-order EWPT may be able to be detected in the $\kappa_0$-$\Lambda$ plane.
The solid and dashed red lines of this figure correspond to the same as the red region of Fig.~\ref{abH3} for $r=1$ and 0.3, respectively. 
In the red region, the fraction of PBHs can be sizable with $f_{\rm PBH}>10^{-4}$ for $0.3<r<1$.
The PBH observation may be able to be used to explore the strongly first-order EWPT in such a wide parameter region.

 \section{Conclusion}

We have investigated the production of PBHs from first-order EWPT in the framework of the naHEFT, in which non-decoupling quantum effects are properly described. 
Since the mass of PBHs from first-order EWPT is about $10^{-5}$ of the solar mass, the current and future microlensing observations such as Subaru HSC, OGLE, PRIME and Roman Space Telescope may be able to probe the EWPT. 
We have examined the parameter region where PBHs can be produced from first-order EWPT and have found that the PBH observations could probe the strongly first-order EWPT of the naHEFT. 
Complementarity in testing the strongly first-order EWPT by future collider experiments, GW and PBH observations has also been investigated. 
From Fig.~\ref{abH4}, the PBH observation may be able to be used to explore wide parameter region. 
Therefore, the PBH observation is a powerful tool to probe the strongly first-order EWPT.


\begin{acknowledgments}

The work of S.~K. was supported by the Grant-in-Aid on Innovative Areas, the Ministry of Education, Culture, Sports, Science and Technology, No.~16H06492, and by the JSPS KAKENHI Grant No.~20H00160.
The work of T.~T. was supported by JSPS KAKENHI Grant Number 19K03874.
The work of M. T. was supported in part by JSPS KAKENHI Grant No. JP21J10645.

\end{acknowledgments}

\appendix

 \section{PBH from first-order phase transition}

We briefly review the mechanism of PBH production from first-order phase transition by the treatment in Ref.~\cite{Liu:2021svg}. 
The critical bubble nucleation rate per unit volume per unit time is given by~\cite{Linde:1981zj}
\begin{equation}
\label{eq:decay}
\Gamma (T) \simeq  T^4 \left(\frac{S_3}{2\pi T}\right)^{\frac{3}{2}} \exp \left(-S_3/T\right) , 
 \end{equation}
 where $S_3$ is the three dimensional Euclidean action. 
  From this equation, the probability of existing the false vacuum is given by~\cite{Turner:1992tz}
\begin{equation}
\label{eq:fraction}
F(t) = \exp\left[ -\frac{4\pi}{3}\int^t_{t_i} dt' \Gamma(t') a^3(t) r^3(t,t') \right],
\end{equation}
where $t_i$ is nucleation time of a first bubble in the Universe, $a(t)$ is the scale factor, and $r(t,t')$ is the comoving radius of the true vacuum from $t'$ to $t$, which is given by
\begin{equation}
r(t,t') \equiv \int^t_{t'}  \frac{1}{ a(\tilde{t})}d\tilde{t}.
\end{equation}
We here assume that the bubble wall velocity is closed to the light speed.
In this case, the bubble wall can be treated as radiation, and then the total radiation energy density is 
\begin{equation}
\rho_R = \rho_r+\rho_w.
\end{equation}
The evolution of this energy density can be described by 
\begin{equation}
\label{evo}
\frac{d\rho_R}{dt}+4H\rho_R = -\frac{d\rho_v}{dt}.
\end{equation}
The vacuum energy density $\rho_v$ is given by 
\begin{equation}
\label{eq:rho_v}
 \rho_v(t)\equiv F(t) \Delta V,
 \end{equation}
where the $ \Delta V$ is the difference in the potential energy density between false and true vacua.
 The scale factor evolution can be determined by the Friedmann equation
\begin{equation}
\label{H}
H^2 =  \left(\frac{1}{ a(t)}\frac{da(t)}{ dt} \right)^2=\frac{1}{3} (\rho_v+\rho_R),
\end{equation}
where we take the unit of $M_{\rm pl}=1$.
The probability that the Hubble volume collapses into a PBH is given by
\begin{equation}
\label{prob}
P(t_n) = \exp\left[ -\frac{4\pi}{3}\int^{t_n}_{t_i} dt \frac{a^3(t)}{a^3(t_{\rm PBH})} \frac{1}{H^3(t_{\rm PBH})} \Gamma(t) \right],
\end{equation}
where $t_n$ is time of nucleation of a bubble in the Hubble volume, and $t_{\rm PBH}$ is the time of production of the PBH.
The production time $t_{\rm PBH}$ can be obtained when the density contrast between the inside and outside of the Hubble volume exceeds the critical value $\delta_c=0.45$~\cite{Musco:2004ak,Harada:2013epa}.
The energy density contrast $\delta$ is given by Eq.~(\ref{cont}).
When $\delta$ exceeds $\delta_c$, 
the inside of the Hubble volume can gravitationally collapse into a PBH.
The PBH mass is given by
\begin{equation}
\label{PBHmass}
M_{\rm PBH} \sim \frac{4\pi}{3}H^{-3} (t_{\rm PBH}) \rho_c = 4\pi H^{-1} (t_{\rm PBH}).
\end{equation}
In the case of the EWPT, the PBH mass is $M_{\rm PBH}^{\rm EW} \sim 10^{-5} M_{\odot}$, where $M_{\odot}$ is the solar mass.
The fraction of the PBH from the first-order EWPT in dark matter density $f_{\rm PBH}$ may be observed by microlensing experiments.
For the EWPT, the fraction $f_{\rm PBH}^{\rm EW}$ is given by
\begin{equation}
\label{PBHfractionEWPT}
f_{\rm PBH}^{\rm EW} \sim1.49 \times 10^{11} \left( \frac{0.25}{\Omega_{\rm CDM}} \right) \left( \frac{T_{\rm PBH}}{100{\rm GeV}} \right) P(t_n),
\end{equation}
where $\Omega_{\rm CDM}$ is the current energy density of cold dark matter normalized by the total energy density and $T_{\rm PBH}$ is the temperature at the PBHs production.
Regions of $f_{\rm PBH}>10^{-2}$ are already in the reach of current observations at Subaru HSC~\cite{Niikura:2017zjd} and OGLE~\cite{Niikura:2019kqi}.
On the other hand, future microlensing experiments, such as Roman Space Telescope, can test the parameter region with the fraction $f_{\rm PBH}>10^{-4}$~\cite{Roman2} .



\begin{thebibliography}{99}


\bibitem{ATLAS:2012yve}
G.~Aad \textit{et al.} [ATLAS],
Phys. Lett. B \textbf{716}, 1-29 (2012).

\bibitem{CMS:2012qbp}
S.~Chatrchyan \textit{et al.} [CMS],
Phys. Lett. B \textbf{716}, 30-61 (2012).


\bibitem{Kuzmin:1985mm} 
  V.~A.~Kuzmin, V.~A.~Rubakov and M.~E.~Shaposhnikov,
  Phys.\ Lett.\ B {\bf 155}, 36 (1985).
  
\bibitem{Sakharov:1967dj} 
  A.~D.~Sakharov,
  Pisma Zh.\ Eksp.\ Teor.\ Fiz.\  {\bf 5}, 32 (1967)
  [JETP Lett.\  {\bf 5}, 24 (1967)]
  [Sov.\ Phys.\ Usp.\  {\bf 34}, 392 (1991)]
  [Usp.\ Fiz.\ Nauk {\bf 161}, 61 (1991)].
  
  

\bibitem{Dine:1992vs}
M.~Dine, R.~G.~Leigh, P.~Huet, A.~D.~Linde and D.~A.~Linde,
Phys. Lett. B \textbf{283} (1992), 319-325.


\bibitem{Kajantie:1995kf}
K.~Kajantie, M.~Laine, K.~Rummukainen and M.~E.~Shaposhnikov,
Nucl. Phys. B \textbf{466} (1996), 189-258.

  
   
  
\bibitem{Grojean:2004xa}
C.~Grojean, G.~Servant and J.~D.~Wells,
Phys. Rev. D \textbf{71} (2005), 036001.



\bibitem{Delaunay:2007wb}
C.~Delaunay, C.~Grojean and J.~D.~Wells,
JHEP \textbf{04}, 029 (2008).


\bibitem{Cai:2017tmh}
R.~G.~Cai, M.~Sasaki and S.~J.~Wang,
JCAP \textbf{08}, 004 (2017).


\bibitem{Croon:2020cgk}
D.~Croon, O.~Gould, P.~Schicho, T.~V.~I.~Tenkanen and G.~White,
JHEP \textbf{04}, 055 (2021).


\bibitem{Hashino:2021qoq}
K.~Hashino, S.~Kanemura and T.~Takahashi,
Phys. Lett. B \textbf{833} (2022), 137261.



\bibitem{Banta:2022rwg}
I.~Banta,
JHEP \textbf{06}, 099 (2022).


\bibitem{Kanemura:2022txx}
S.~Kanemura, R.~Nagai and M.~Tanaka,
JHEP \textbf{06} (2022), 027.



  
  


\bibitem{Turok:1991uc}
N.~Turok and J.~Zadrozny,
Nucl. Phys. B \textbf{369}, 729 (1992).

\bibitem{Espinosa:1993bs}
J.~R.~Espinosa and M.~Quiros,
Phys. Lett. B \textbf{305}, 98 (1993).

\bibitem{Funakubo:1993jg} 
  K.~Funakubo, A.~Kakuto and K.~Takenaga,
  Prog.\ Theor.\ Phys.\  {\bf 91}, 341 (1994). 

\bibitem{Cottingham:1995cj}
W.~N.~Cottingham and N.~Hasan,
Phys. Rev. D \textbf{51}, 866 (1995).

\bibitem{Cline:1996mga}
J.~M.~Cline and P.~A.~Lemieux,
Phys. Rev. D \textbf{55}, 3873 (1997).

\bibitem{Moreno:1996zm}
J.~M.~Moreno, D.~H.~Oaknin and M.~Quiros,
Nucl. Phys. B \textbf{483}, 267 (1997).


\bibitem{Kanemura:2004ch} 
  S.~Kanemura, Y.~Okada and E.~Senaha,
  Phys.\ Lett.\ B {\bf 606}, 361 (2005).
  
\bibitem{Funakubo:2005pu} 
  K.~Funakubo, S.~Tao and F.~Toyoda,
  Prog.\ Theor.\ Phys.\  {\bf 114}, 369 (2005). 

\bibitem{Fromme:2006cm}
L.~Fromme, S.~J.~Huber and M.~Seniuch,
JHEP \textbf{11}, 038 (2006).

\bibitem{Profumo:2007wc} 
  S.~Profumo, M.~J.~Ramsey-Musolf and G.~Shaughnessy,
  JHEP {\bf 0708}, 010 (2007). 

\bibitem{Ahriche:2007jp}
A.~Ahriche,
Phys. Rev. D \textbf{75}, 083522 (2007).

\bibitem{Noble:2007kk} 
  A.~Noble and M.~Perelstein,
  Phys.\ Rev.\ D {\bf 78}, 063518 (2008). 
  
\bibitem{Aoki:2008av} 
  M.~Aoki, S.~Kanemura and O.~Seto,
  Phys.\ Rev.\ Lett.\  {\bf 102}, 051805 (2009). 

\bibitem{Funakubo:2009eg}
K.~Funakubo and E.~Senaha,
Phys. Rev. D \textbf{79}, 115024 (2009).

\bibitem{Kanemura:2011fy} 
  S.~Kanemura, E.~Senaha and T.~Shindou,
  Phys.\ Lett.\ B {\bf 706}, 40 (2011). 

\bibitem{Espinosa:2011ax}
J.~R.~Espinosa, T.~Konstandin and F.~Riva,
Nucl. Phys. B \textbf{854}, 592 (2012).

\bibitem{Gil:2012ya} 
  G.~Gil, P.~Chankowski and M.~Krawczyk,
  Phys.\ Lett.\ B {\bf 717}, 396 (2012). 

\bibitem{Fuyuto:2014yia}
K.~Fuyuto and E.~Senaha,
Phys. Rev. D \textbf{90}, 015015 (2014).

\bibitem{Tamarit:2014dua} 
  C.~Tamarit,
  Phys.\ Rev.\ D {\bf 90}, no. 5, 055024 (2014). 
  
\bibitem{Kanemura:2014cka} 
  S.~Kanemura, N.~Machida and T.~Shindou,
  Phys.\ Lett.\ B {\bf 738}, 178 (2014). 
  
\bibitem{Profumo:2014opa} 
  S.~Profumo, M.~J.~Ramsey-Musolf, C.~L.~Wainwright and P.~Winslow,
  Phys.\ Rev.\ D {\bf 91}, no. 3, 035018 (2015). 
  
\bibitem{Blinov:2015vma} 
  N.~Blinov, S.~Profumo and T.~Stefaniak,
  JCAP {\bf 1507}, no. 07, 028 (2015). 
  
\bibitem{Fuyuto:2015vna} 
  K.~Fuyuto and E.~Senaha,
  Phys.\ Lett.\ B {\bf 747}, 152 (2015).

\bibitem{Karam:2015jta} 
  A.~Karam and K.~Tamvakis,
  Phys.\ Rev.\ D {\bf 92}, no. 7, 075010 (2015).   

\bibitem{Basler:2016obg}
P.~Basler, M.~Krause, M.~Muhlleitner, J.~Wittbrodt and A.~Wlotzka,
JHEP \textbf{02}, 121 (2017).

\bibitem{Dorsch:2017nza}
G.~C.~Dorsch, S.~J.~Huber, K.~Mimasu and J.~M.~No,
JHEP \textbf{12}, 086 (2017).

\bibitem{Ghorbani:2017jls}
P.~H.~Ghorbani,
JHEP \textbf{08}, 058 (2017).

\bibitem{Bernon:2017jgv}
J.~Bernon, L.~Bian and Y.~Jiang,
JHEP \textbf{05}, 151 (2018).

\bibitem{Chiang:2017nmu}
C.~W.~Chiang, M.~J.~Ramsey-Musolf and E.~Senaha,
Phys. Rev. D \textbf{97}, 015005 (2018).

\bibitem{Ghorbani:2019itr}
K.~Ghorbani and P.~H.~Ghorbani,
JHEP \textbf{12}, 077 (2019).

\bibitem{Barman:2019oda}
B.~Barman, A.~Dutta Banik and A.~Paul,
Phys. Rev. D \textbf{101}, 055028 (2020).





\bibitem{Enomoto:2021dkl}
K.~Enomoto, S.~Kanemura and Y.~Mura,
JHEP \textbf{01}, 104 (2022).



\bibitem{Kanemura:2022ozv}
S.~Kanemura and M.~Tanaka,
Phys. Rev. D \textbf{106}, 035012 (2022).

  
  

\bibitem{Cepeda:2019klc}
M.~Cepeda, S.~Gori, P.~Ilten, M.~Kado, F.~Riva, R.~Abdul Khalek, A.~Aboubrahim, J.~Alimena, S.~Alioli and A.~Alves, \textit{et al.}
CERN Yellow Rep. Monogr. \textbf{7} (2019), 221-584.



\bibitem{Asner:2013psa} 
  D.~M.~Asner {\it et al.},
  arXiv:1310.0763 [hep-ph].

\bibitem{Moortgat-Picka:2015yla} 
  G.~Moortgat-Pick {\it et al.},
  Eur.\ Phys.\ J.\ C {\bf 75}, no. 8, 371 (2015).

\bibitem{Fujii:2015jha} 
  K.~Fujii {\it et al.},
  arXiv:1506.05992 [hep-ex].




\bibitem{Grojean:2006bp}
C.~Grojean and G.~Servant,
Phys. Rev. D \textbf{75} (2007), 043507.




\bibitem{Klein:2015hvg} 
  A.~Klein {\it et al.},
  Phys.\ Rev.\ D {\bf 93}, 024003 (2016).



\bibitem{Yagi:2011wg} 
  K.~Yagi and N.~Seto,
  Phys.\ Rev.\ D {\bf 83}, 044011 (2011)
  Erratum: [Phys.\ Rev.\ D {\bf 95}, 109901 (2017)].



\bibitem{Kakizaki:2015wua}
M.~Kakizaki, S.~Kanemura and T.~Matsui,
Phys. Rev. D \textbf{92} (2015), 115007.

\bibitem{Hashino:2016rvx}
K.~Hashino, M.~Kakizaki, S.~Kanemura and T.~Matsui,
Phys. Rev. D \textbf{94} (2016), 015005.


\bibitem{Kobakhidze:2016mch}
A.~Kobakhidze, A.~Manning and J.~Yue,
Int. J. Mod. Phys. D \textbf{26} (2017) no.10, 1750114.


\bibitem{Huang:2016cjm}
P.~Huang, A.~J.~Long and L.~T.~Wang,
Phys. Rev. D \textbf{94} (2016), 075008.

\bibitem{Hashino:2016xoj}
K.~Hashino, M.~Kakizaki, S.~Kanemura, P.~Ko and T.~Matsui,
Phys. Lett. B \textbf{766} (2017), 49-54.


\bibitem{Artymowski:2016tme}
M.~Artymowski, M.~Lewicki and J.~D.~Wells,
JHEP \textbf{03} (2017), 066.

\bibitem{Beniwal:2017eik}
A.~Beniwal, M.~Lewicki, J.~D.~Wells, M.~White and A.~G.~Williams,
JHEP \textbf{08} (2017), 108.


\bibitem{Huang:2017rzf}
F.~P.~Huang and J.~H.~Yu,
Phys. Rev. D \textbf{98} (2018), 095022.


\bibitem{Hashino:2018zsi}
K.~Hashino, M.~Kakizaki, S.~Kanemura, P.~Ko and T.~Matsui,
JHEP \textbf{06} (2018), 088.


\bibitem{Chala:2018ari}
M.~Chala, C.~Krause and G.~Nardini,
JHEP \textbf{07} (2018), 062.


\bibitem{Huang:2018aja}
F.~P.~Huang, Z.~Qian and M.~Zhang,
Phys. Rev. D \textbf{98} (2018), 015014.

\bibitem{Bruggisser:2018mrt}
S.~Bruggisser, B.~Von Harling, O.~Matsedonskyi and G.~Servant,
JHEP \textbf{12} (2018), 099.


\bibitem{Alves:2018oct}
A.~Alves, T.~Ghosh, H.~K.~Guo and K.~Sinha,
JHEP \textbf{12} (2018), 070.

\bibitem{Hashino:2018wee}
K.~Hashino, R.~Jinno, M.~Kakizaki, S.~Kanemura, T.~Takahashi and M.~Takimoto,
Phys. Rev. D \textbf{99} (2019), 075011.

\bibitem{Ahriche:2018rao}
A.~Ahriche, K.~Hashino, S.~Kanemura and S.~Nasri,
Phys. Lett. B \textbf{789} (2019), 119-126.



\bibitem{Chala:2018opy}
M.~Chala, M.~Ramos and M.~Spannowsky,
Eur. Phys. J. C \textbf{79} (2019) no.2, 156.


\bibitem{Alves:2018jsw}
A.~Alves, T.~Ghosh, H.~K.~Guo, K.~Sinha and D.~Vagie,
JHEP \textbf{04} (2019), 052.


\bibitem{Alves:2019igs}
A.~Alves, D.~Gon\c{c}alves, T.~Ghosh, H.~K.~Guo and K.~Sinha,
JHEP \textbf{03} (2020), 053.

\bibitem{Chen:2019ebq}
N.~Chen, T.~Li, Y.~Wu and L.~Bian,
Phys. Rev. D \textbf{101} (2020), 075047.



\bibitem{Liu:2021svg}
J.~Liu, L.~Bian, R.~G.~Cai, Z.~K.~Guo and S.~J.~Wang,
Phys. Rev. D \textbf{105} (2022), 2.


\bibitem{Hawking:1971ei}
S.~Hawking,
Mon. Not. Roy. Astron. Soc. \textbf{152} (1971), 75.

\bibitem{Carr:1974nx}
B.~J.~Carr and S.~W.~Hawking,
Mon. Not. Roy. Astron. Soc. \textbf{168} (1974), 399-415.

\bibitem{Carr:1975qj}
B.~J.~Carr,
Astrophys. J. \textbf{201} (1975), 1-19.

\bibitem{Kodama:1982sf}
H.~Kodama, M.~Sasaki and K.~Sato,
Prog. Theor. Phys. \textbf{68} (1982), 1979.

\bibitem{Hawking:1982ga}
S.~W.~Hawking, I.~G.~Moss and J.~M.~Stewart,
Phys. Rev. D \textbf{26} (1982), 2681.




\bibitem{HSC}
{\tt https://hsc.mtk.nao.ac.jp/ssp/}

\bibitem{OGLE}
{\tt http://ogle.astrouw.edu.pl}

\bibitem{PRIME}
{\tt http://www-ir.ess.sci.osaka-u.ac.jp/prime/index.html}


\bibitem{Roman}
{\tt https://roman.gsfc.nasa.gov}




\bibitem{Postma:2020toi}
M.~Postma and G.~White,
JHEP \textbf{03} (2021), 280.





\bibitem{Kanemura:2021fvp}
S.~Kanemura and R.~Nagai,
JHEP \textbf{03} (2022), 194.

\bibitem{Coleman:1973jx}
S.~R.~Coleman and E.~J.~Weinberg,
Phys. Rev. D \textbf{7} (1973), 1888-1910.


\bibitem{Linde:1981zj}
A.~D.~Linde,
Nucl. Phys. B \textbf{216}, 421 (1983).
[erratum: Nucl. Phys. B \textbf{223}, 544 (1983).]

\bibitem{Kanemura:2004mg}
S.~Kanemura, Y.~Okada, E.~Senaha and C.~P.~Yuan,
Phys. Rev. D \textbf{70} (2004), 115002.


\bibitem{Aoki:2012jj}
M.~Aoki, S.~Kanemura, M.~Kikuchi and K.~Yagyu,
Phys. Rev. D \textbf{87} (2013), 015012.

\bibitem{Arhrib:2015hoa}
A.~Arhrib, R.~Benbrik, J.~El Falaki and A.~Jueid,
JHEP \textbf{12} (2015), 007.


\bibitem{Hashino:2015nxa}
K.~Hashino, S.~Kanemura and Y.~Orikasa,
Phys. Lett. B \textbf{752} (2016), 217-220.


\bibitem{Kanemura:2016lkz}
S.~Kanemura, M.~Kikuchi and K.~Yagyu,
Nucl. Phys. B \textbf{917} (2017), 154-177.



\bibitem{Braathen:2019pxr}
J.~Braathen and S.~Kanemura,
Phys. Lett. B \textbf{796} (2019), 38-46.

\bibitem{Braathen:2019zoh}
J.~Braathen and S.~Kanemura,
Eur. Phys. J. C \textbf{80} (2020) no.3, 227.

\bibitem{Braathen:2020vwo}
J.~Braathen, S.~Kanemura and M.~Shimoda,
JHEP \textbf{03} (2021), 297.


\bibitem{Bambade:2019fyw}
P.~Bambade, T.~Barklow, T.~Behnke, M.~Berggren, J.~Brau, P.~Burrows, D.~Denisov, A.~Faus-Golfe, B.~Foster and K.~Fujii, \textit{et al.}
[arXiv:1903.01629 [hep-ex]].





\bibitem{Caprini:2015zlo}
C.~Caprini, M.~Hindmarsh, S.~Huber, T.~Konstandin, J.~Kozaczuk, G.~Nardini, J.~M.~No, A.~Petiteau, P.~Schwaller and G.~Servant, \textit{et al.}
JCAP \textbf{04}, 001 (2016).

\bibitem{Espinosa:2010hh}
J.~R.~Espinosa, T.~Konstandin, J.~M.~No and G.~Servant,
JCAP \textbf{06}, 028 (2010).




\bibitem{Cline:2021iff}
J.~M.~Cline, A.~Friedlander, D.~M.~He, K.~Kainulainen, B.~Laurent and D.~Tucker-Smith,
Phys. Rev. D \textbf{103} (2021), 123529.



\bibitem{Turner:1992tz}
M.~S.~Turner, E.~J.~Weinberg and L.~M.~Widrow,
Phys. Rev. D \textbf{46} (1992), 2384-2403.



\bibitem{Musco:2004ak}
I.~Musco, J.~C.~Miller and L.~Rezzolla,
Class. Quant. Grav. \textbf{22} (2005), 1405-1424.

\bibitem{Harada:2013epa}
T.~Harada, C.~M.~Yoo and K.~Kohri,
Phys. Rev. D \textbf{88} (2013), 084051.
[erratum: Phys. Rev. D \textbf{89} (2014), 029903.]




\bibitem{Niikura:2017zjd}
H.~Niikura, M.~Takada, N.~Yasuda, R.~H.~Lupton, T.~Sumi, S.~More, T.~Kurita, S.~Sugiyama, A.~More and M.~Oguri, \textit{et al.}
Nature Astron. \textbf{3} (2019) no.6, 524-534.

\bibitem{Niikura:2019kqi}
H.~Niikura, M.~Takada, S.~Yokoyama, T.~Sumi and S.~Masaki,
Phys. Rev. D \textbf{99} (2019), 083503.


\bibitem{Roman2} 
S.~A. Johnson, M. Penny, B.~S. Gaudi, et al. 2020, Astron. J., {\bf 160}, 123 (2020). 



  \end{thebibliography}
\end{document}